\documentclass[preprint,3p]{elsarticle}

\usepackage{amssymb}
\usepackage{amsthm}
\usepackage[mathlines]{lineno}
\usepackage{graphicx}
\usepackage{url}
\usepackage{amsmath}
\usepackage{amsfonts}
\usepackage{bm}
\usepackage{textcomp}
\usepackage{subfigure}
\usepackage{multicol}
\usepackage{verbatim}
\usepackage{rotating}
\usepackage[colorlinks,linkcolor=red,citecolor=red]{hyperref}
\usepackage{float}
\usepackage{epsfig}
\usepackage{dcolumn}
\usepackage{bm}
\usepackage{color}
\usepackage{pstricks}
\usepackage{pst-node}
\usepackage{times}
\usepackage{indentfirst}
\usepackage[english]{babel}
\usepackage{siunitx}
\usepackage{etoolbox}
\journal{Physics Letters B}
\makeatletter
\newcommand{\psip}{\psi(3686)}
\newcommand{\jpsi}{J/\psi}
\newcommand{\etap}{\eta^{\prime}}
\newcommand{\bcl}{\begin{center}}
\newcommand{\ecl}{\end{center}}
\usepackage{setspace}

\begin{document}
\begin{frontmatter}
\title{ {\bf \boldmath Observation of $\psi(3686) \to \etap e^+ e^- $}}
\author{
\begin{small}
\begin{center}
M.~Ablikim$^{1}$, M.~N.~Achasov$^{9,d}$, S.~Ahmed$^{14}$, M.~Albrecht$^{4}$, M.~Alekseev$^{55A,55C}$, A.~Amoroso$^{55A,55C}$, F.~F.~An$^{1}$, Q.~An$^{52,42}$, J.~Z.~Bai$^{1}$, Y.~Bai$^{41}$
% ...
\\
\vspace{0.2cm}
(BESIII Collaboration)\\
\vspace{0.2cm} {\it
$^{1}$ Institute of High Energy Physics, Beijing 100049, People's Republic of China\\
% ...
$^{4}$ Bochum Ruhr-University, D-44780 Bochum, Germany\\
% ...
$^{9}$ G.I. Budker Institute of Nuclear Physics SB RAS (BINP), Novosibirsk 630090, Russia\\
% ...
$^{14}$ Helmholtz Institute Mainz, Johann-Joachim-Becher-Weg 45, D-55099 Mainz, Germany\\
% ...
$^{41}$ Southeast University, Nanjing 211100, People's Republic of China\\
$^{42}$ State Key Laboratory of Particle Detection and Electronics, Beijing 100049, Hefei 230026, People's Republic of China\\
% ...
$^{52}$ University of Science and Technology of China, Hefei 230026, People's Republic of China\\
% ...
$^{55}$ (A)University of Turin, I-10125, Turin, Italy; (B)University of Eastern Piedmont, I-15121, Alessandria, Italy; (C)INFN, I-10125, Turin, Italy\\
% ...
$^{d}$ Also at the Novosibirsk State University, Novosibirsk, 630090, Russia\\
% ...
}\end{center}
\vspace{0.4cm}
\end{small}
}

\begin{abstract}
Using a data sample of $448.1 \times 10^6$ $\psi(3686)$ events collected with the BESIII detector at the BEPCII collider, we report the first observation of the electromagnetic Dalitz decay $\psi(3686) \to \eta' e^+ e^-$, with significances of 7.0$\sigma$ and 6.3$\sigma$ when reconstructing the $\etap$ meson via its decay modes $\eta'\to\gamma \pi^+ \pi^-$ and $\eta'\to\pi^+\pi^-\eta$ ($\eta \to \gamma\gamma$), respectively. The weighted average branching fraction is determined to be  $\mathcal{B}(\psi(3686) \to \eta' e^+ e^-)= (1.90 \pm 0.25 \pm 0.11) \times 10^{-6}$, where the first uncertainty is statistical and the second systematic.
\\
\end{abstract}
\end{frontmatter}

\begin{multicols}{2}
\section{\boldmath Introduction}
The electromagnetic (EM) Dalitz decays $V \to P\ell^+\ell^-$, where $V$ is a vector meson ($V = \rho, \omega, \phi, \psi $), $P$  a pseudoscalar meson ($P = \pi^0, \eta, \eta'$) and $\ell$ a lepton ($\ell= e , \mu$), is of great interest for our understanding of both the intrinsic structure of hadrons and the fundamental mechanisms of the interactions between  photons and hadrons~\cite{lansberg}. These Dalitz decays proceed via a two-body radiative process of $V$ decaying into $P$ and an  off-shell photon, from which the lepton pair in the final state originates. The universal decay width of these Dalitz decays can be normalized to that of the corresponding radiative process $V\to P\gamma$ and can be parameterized as a product of  the quantum electrodynamics prediction for a point-like particle  and the transition form factor (TFF) $F(q^2)$ at the $V$-$P$ transition vertex~\cite{lansberg}, where $q^2=M^2_{\ell^+\ell^-}c^2$ is the four-momentum transfer squared. Knowledge of  the $q^2$-dependent TFF thus provides information about the EM structure arising at the $V$-$P$ vertex.

EM Dalitz decays have been widely observed for light unflavored mesons, such as $\omega \to \pi^0 e^+e^-$~\cite{cmd2_omegapi0ee,a2_omegapi0ee}, $\omega \to \pi^0 \mu^+\mu^-$~\cite{na60_omegapi0mumu}, $\phi \to \pi^0e^+e^-$~\cite{kloe_phipi0ee} and $\phi \to \eta e^+e^-$~\cite{snd_phietaee,kloe_phietaee}. The investigation of these  decays  motivated the authors of Ref.~\cite{fujinlin_modphya} to study the  charmonium decays $\jpsi \to P \ell^+\ell^-$ and to calculate  the branching fractions based on a monopole TFF  $F(q^2)=1/(1-q^2/\Lambda^2)$  using a vector meson dominance  model. Here $\Lambda$ is an effective pole mass accounting for the  overall effects from all possible resonance poles and scattering terms in the time-like kinematic region.  The  charmonium EM Dalitz decays $\jpsi \to Pe^+e^-$ have been previously observed by the BESIII experiment  using a data sample of $2.25 \times 10^8$ $\jpsi$ events~\cite{chuxkpee}. The results agree well with the theoretical predictions~\cite{fujinlin_modphya} for the $P = \eta, \etap$ cases. However,  similar EM Dalitz decays have never been studied in $\psip$ decays. The investigation  of such  processes will be important to understand the interaction of charmonium vector states with photons, and helpful for further studies on the $\psi \to VP$ process, including the related $\rho\pi$ puzzle~\cite{zhaoq_VP}. In this Letter, we report the first observation of the charmonium EM Dalitz decay $\psip \to \etap e^+e^-$ using a data sample of $448.1 \times 10^6$ $\psip$ events ($107.0 \times 10^6$~\cite{tnum09} in 2009 and $341.1 \times 10^6$~\cite{tnum}  in 2012) collected with the BESIII detector~\cite{bes3_design}. Here, the intermediate $\etap$ meson is reconstructed via two decay modes,  $\etap \to \gamma \pi^+\pi^-$ and $\etap \to \pi^+\pi^- \eta $ with $\eta \to \gamma\gamma$.

\section{\boldmath The BESIII experiment and Monte Carlo simulation}
The BESIII detector~\cite{bes3_design} is a magnetic spectrometer operating at BEPCII, a double ring $e^+e^-$ collider running at center-of-mass (c.m.) energies between  2.0 and 4.6 GeV with a peak luminosity  of $1\times 10^{33}{\rm~cm}^{-2}{\rm s}^{-1}$  at a c.m.\ energy of 3.773 GeV. The cylindrical core of the BESIII detector comprises a helium-gas-based main drift chamber (MDC) to measure the momentum and the ionization energy loss (${\rm d}E/{\rm d}x$) of charged particles, a plastic scintillator time-of-flight (TOF) system for particle identification (PID) information, a CsI(Tl) electromagnetic calorimeter (EMC) to measure photon and electron energies and a multilayer resistive plate chamber muon counter system (MUC) to identify muons. The MDC, TOF and EMC are enclosed in a superconducting solenoidal magnet providing a 1.0~T magnetic field. The geometrical acceptance is 93\% of  $4\pi$  for charged particles and photons. The momentum resolution  is 0.5\% for charged particles with transverse momentum of 1~GeV/$c$, and the energy resolution for photons  is 2.5\% (5\%) at 1~GeV in the barrel (end cap) EMC.

Monte Carlo (MC) simulations are used to optimize the event selection criteria, to investigate  potential backgrounds and to determine the detection efficiency. The {\sc geant4}-based~\cite{geant4} simulation  includes the description of geometry and  material of the BESIII detector, the detector response, digitization models and  tracking of the detector  running conditions and its performance. An inclusive  MC sample containing $506 \times 10^6$ generic $\psip$ decays is used to study the potential backgrounds. The production of the $\psip$ resonance is simulated by the MC  generator {\sc kkmc}~\cite{kkmc}, in which the beam energy spread and initial state radiation (ISR) effects are also included. The known decay modes of $\psip$ are generated by {\sc evtgen}~\cite{evtgen} with branching fractions taken from the  Particle Data Group (PDG)~\cite{pdg16}, while the remaining unknown decay modes are generated according to the {\sc lundcharm}~\cite{lunccharm} model. When generating the process $\psip \to \etap e^+e^-$, the TFF is parameterized as a monopole form factor with $\Lambda = 3.773$ GeV/$c^2$. For the decay of $\etap \to \gamma \pi^+\pi^-$, the generator takes into account the $\rho$-$\omega$ interference and box anomaly~\cite{qinlq}. The decays of $\etap \to \pi^+\pi^-\eta$ and $\eta \to \gamma \gamma$ are generated with a phase space model. The analysis is performed in the framework of the BESIII offline software system which takes care of the detector calibration and  event reconstruction.

\section{\boldmath Data analysis}

Charged tracks in  BESIII  are reconstructed from ionization signals of particles in the MDC. The point of closest approach of every charged track to the $e^+e^-$ interaction point (IP) is required to be within  $\pm 10$ cm in the beam direction and within 1 cm in the plane perpendicular to the beam direction. The polar angle $\theta$ between the direction of a charged track and that of the beam must satisfy  $|\cos \theta| <  0.93$ for an effective measurement in the MDC. Four charged tracks are required with zero net charge for each candidate event. The combined information of the energy loss ${\rm d}E/{\rm d}x$ and TOF is used to calculate PID confidence levels (C.L.) for the electron, pion and kaon hypotheses. Both the electron and positron require the highest PID C.L.\ for the electron hypothesis while the other two charged tracks are assumed to be pion candidates without any PID requirements.

Electromagnetic showers are reconstructed from clusters of energy depositions in the EMC. The shower energy of photon candidates in the EMC should be greater than 25 MeV in the barrel region ($|\cos \theta| < 0.80$) or 50 MeV in the endcap region ($0.86<|\cos \theta| < 0.92$), whereas the showers located in the transition regions between the barrel and the endcaps are excluded due to bad reconstruction. The photon candidates  are required to be separated from the extrapolated positions of any charged track by more than $10^{\circ}$ to exclude  showers from charged particles. To suppress electronic noise and energy deposition unrelated to the event, the time at which the photon is recorded in the EMC with respect to the collision must be less than 700 ns. We require at least one photon  in the decay mode $\etap \to \gamma \pi^+\pi^-$ and at least two photons for the decay $\etap \to \pi^+\pi^-\eta$.

A vertex constraint is enforced on the four charged tracks $\pi^+\pi^-e^+e^-$  to ensure they originate from the IP. To improve the resolution and suppress backgrounds, a kinematic fit with an energy-momentum constraint (4C) is performed. For events with more than the required number of photons, only the combination with the least $\chi^2_{\rm 4C}$ is retained. In all cases, events with $\chi^2_{\rm 4C}<80$ are kept for further analysis.

The dominant background originates from the decay of  $\psip \to \pi^+\pi^- J/\psi, J/\psi \to \ell^+\ell^-(\gamma)$ due to the sizable branching fraction ($34.49\pm0.30)$\%~\cite{pdg16} of the decay $\psip \to \pi^+\pi^- \jpsi$. For the $\etap \to \gamma \pi^+\pi^-$ mode, to suppress the huge background from $\psip \to \pi^+\pi^- \jpsi, \jpsi \to e^+e^-$ we require the recoil mass of the $\pi^+\pi^-$ pair  $RM(\pi^+\pi^-)$ to be smaller than 2.9 GeV/$c^2$, with which about 99.8\% of the  background events  are  removed. Events of the type $\psip \to \pi^+\pi^-\jpsi,\jpsi \to \mu^+\mu^-$ survive the selection when $\pi$ or $\mu$ candidates are misidentified as electrons. An additional  criterion  $E/p > 0.8$ is applied to the track with larger momentum in the $e^+e^-$ pair to further improve the electron identification, where $E$ and $p$ refer to the energy deposition in the EMC  and  momentum measured with the MDC, respectively. The relative selection efficiency of this $E/p$ criterion is more than 98\%.  For the  $\etap \to \pi^+\pi^-\eta$ decay mode,  the background is much lower. The candidate events must satisfy  $RM(\pi^+\pi^-)< 3.2 {\rm ~GeV}/c^2$ to suppress the background from $\psip \to \eta \jpsi,  \jpsi \to e^+e^-, \eta \to \pi^+\pi^-\pi^0, \pi^0 \to \gamma \gamma$, and the invariant mass of the photon pair $M(\gamma\gamma)$ is required to be within the $\eta$ mass window [0.520, 0.575]~GeV/$c^2$.

\begin{figure*}[htb]
\centering
\subfigure{\includegraphics[width=0.24\textwidth]{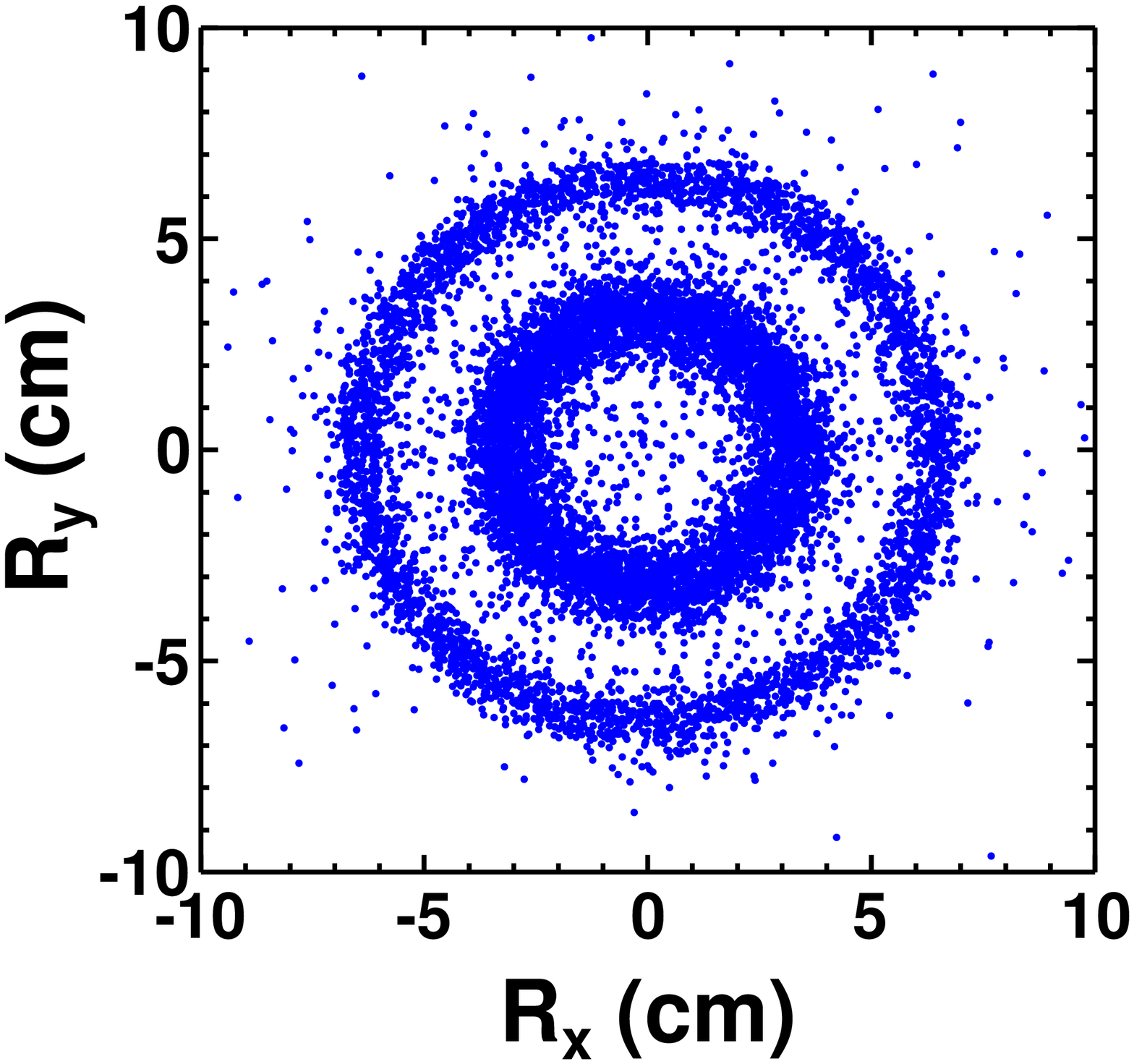}}\put(-23,78){\bf  ~(a)}
\subfigure{\includegraphics[width=0.24\textwidth]{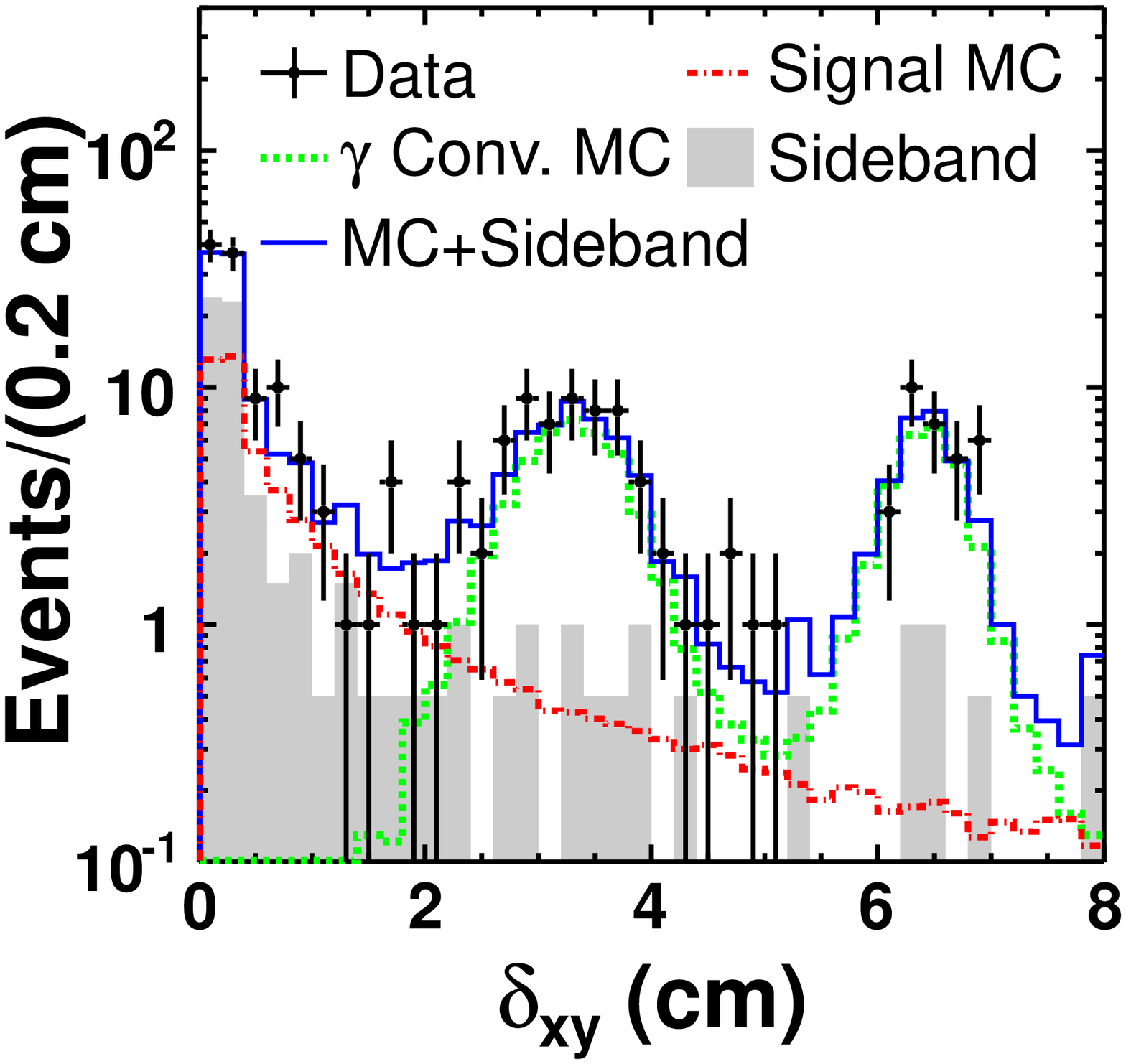}}\put(-23,78){\bf  ~(b)}
\caption{(Color online.) $e^+e^-$  pair vertex position distribution: (a) Scatter plot of $R_y$ versus $R_x$ for simulated MC events of $\psip \to \eta' \gamma, \eta' \to \gamma \pi^+\pi^-$. (b) Distribution of $\delta_{xy}$ in the $\etap \to \gamma \pi^+\pi^-$ mode. The black dots with error bars represent data,  the red dot-dashed  and green dashed histograms show the signal MC simulation and $\gamma$ conversion MC simulation, respectively, the gray shaded histogram shows the non-peaking background estimated from  $\etap$ sideband and  the blue solid histogram is the sum of MC simulations and $\etap$ sideband.}
\label{f_rxy}
\end{figure*}

The radiative decay $\psip\to \eta' \gamma$ contributes as a peaking background to the distributions of the $\gamma \pi^+\pi^-$ and $\gamma \gamma \pi^+\pi^-$ invariant masses ($M(\gamma\pi^+\pi^-)$ and M($\gamma \gamma \pi^+\pi^-)$), if the photon subsequently converts into an $e^+e^-$ pair in  the beam pipe or the inner wall of the MDC. The distance $\delta_{xy}$ from the reconstructed vertex of the $e^+e^-$ pair to the IP in the plane transverse to the beam axis (the $x$-$y$ plane) is used to distinguish such $\gamma$ conversion events from signal events~\cite{xzr_cpc}, where $\delta_{xy} =\sqrt{R_{x}^2+R_{y}^2}$ and $R_{x}$ and $R_{y}$ refer to the coordinates of the reconstructed vertex position in the $x$ and $y$ directions. The scatter plot of $R_y$ versus $R_x$ from a simulated $\gamma$ conversion MC sample $\psip \to \eta' \gamma, \eta' \to \gamma \pi^+\pi^-$ is shown in Fig.~\ref{f_rxy}~(a), where the inner and outer circles refer to the $\gamma$ conversion occurs in  the beam pipe and inner wall of the MDC, respectively. The distributions of $\delta_{xy}$ for the data, $\gamma$ conversion background, and signal from MC simulation are shown in Fig.~\ref{f_rxy}~(b), where the two peaks around $\delta_{xy}=3$ and $\delta_{xy}=6.5$~cm match the positions of the beam pipe and inner wall of the MDC. From the MC study, requiring $\delta_{xy} < 2$~cm will remove more than 97\% of the $\gamma$ conversion background, and the number of remaining events is estimated to be $1.19\pm0.06$ and $0.43\pm0.02$ in the $\etap \to \gamma \pi^+\pi^-$  and $\etap \to \pi^+\pi^-\eta$  mode, respectively.

\begin{figure*}[htb]
\centering
\subfigure{
\includegraphics[width=0.24\textwidth]{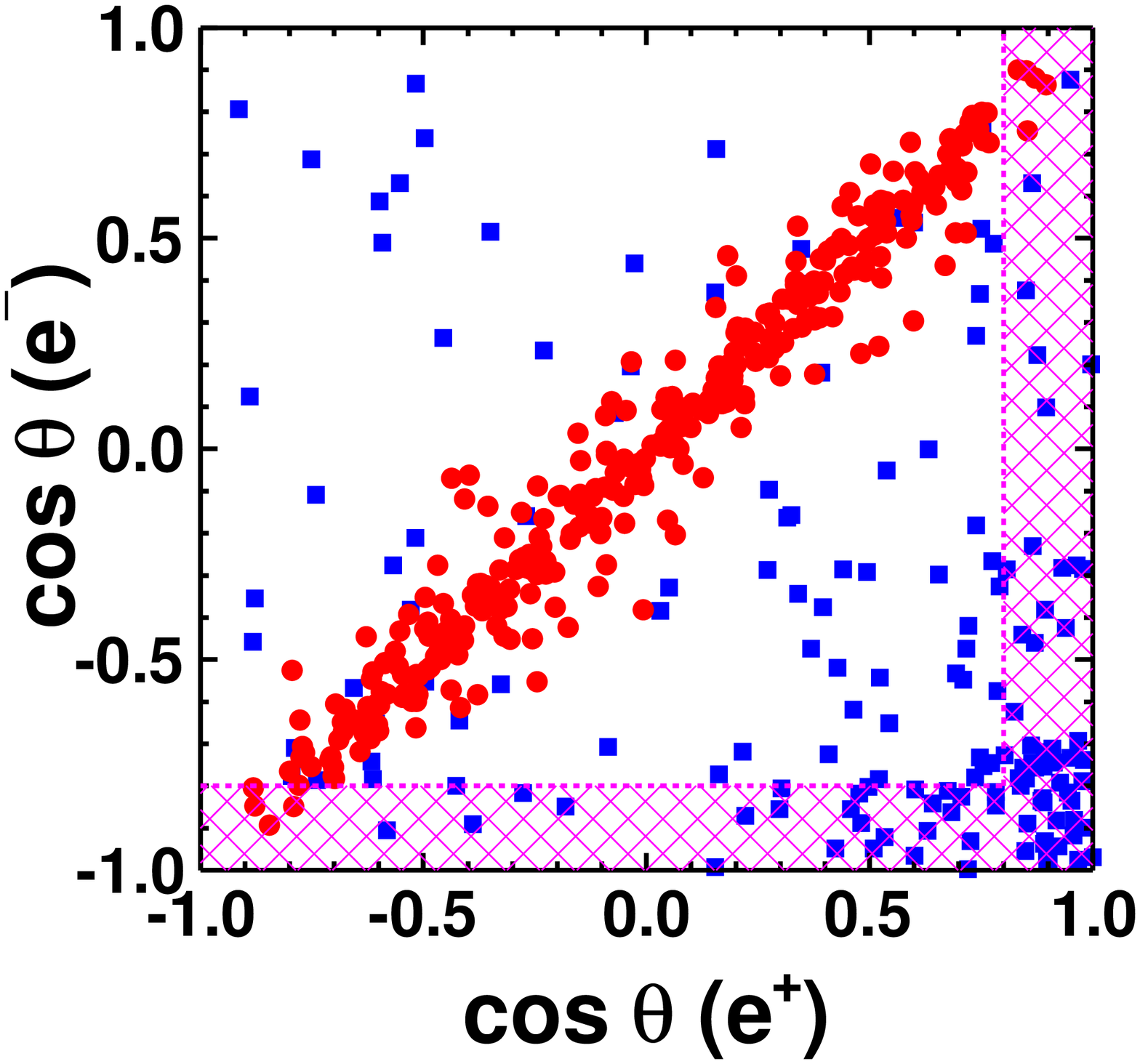}}\put(-90,90){\bf  ~(a)}
\subfigure{
\includegraphics[width=0.24\textwidth]{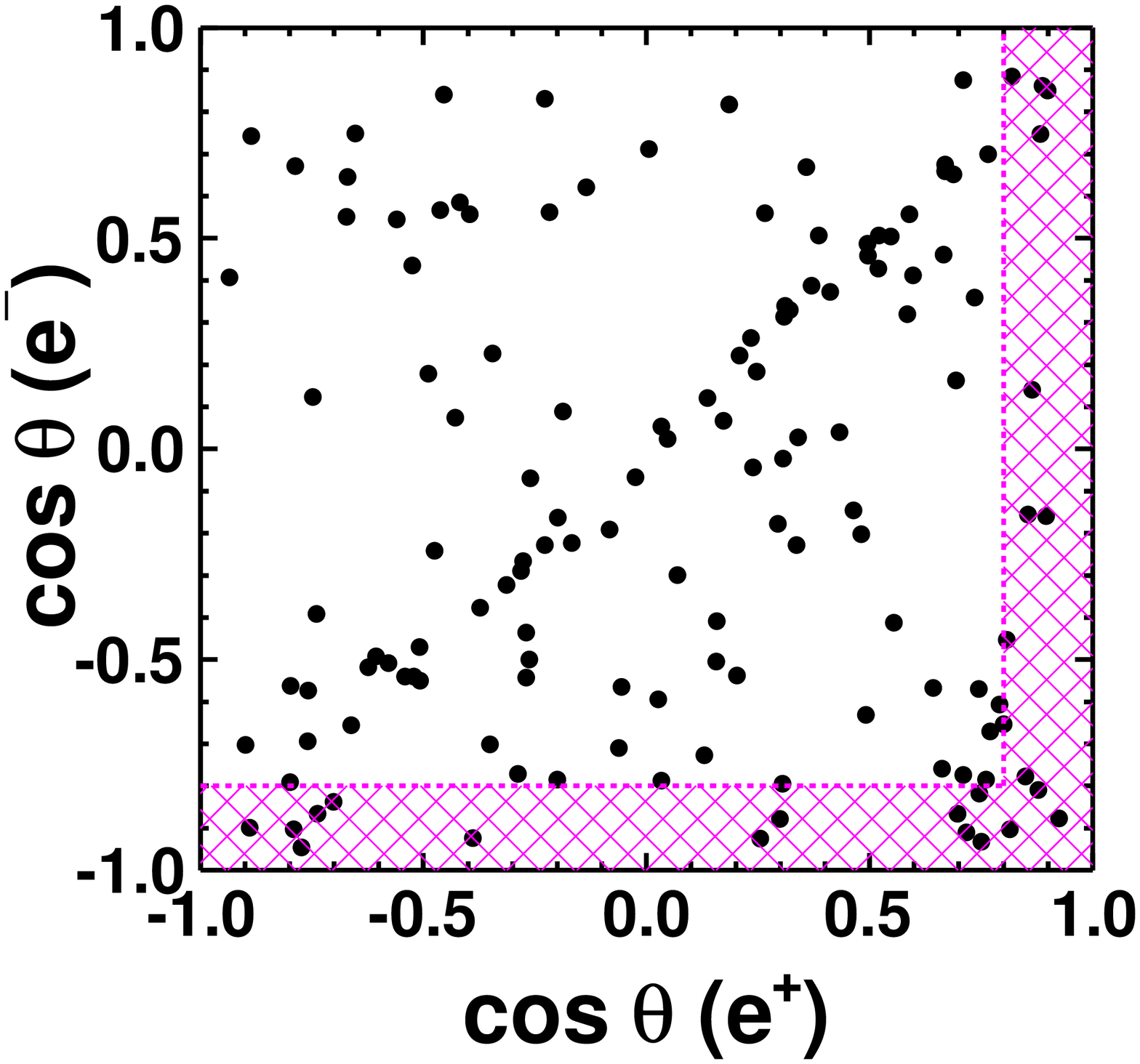}}\put(-90,90){\bf   ~(b)}
\caption{ (Color online.) Scatter plot of  polar angle  $\cos\theta (e^-)$  versus  $\cos\theta (e^+)$.  The areas with pink crosshatched lines refer to the rejected region $\cos \theta(e^+) >0.8$ or $\cos \theta(e^-) < -0.8$. (a)  The red dots represent  signal MC events $\psip \to \etap e^+e^-$ and the blue squares are from $\psi(3770)$ data.  (b) The black dots represent $\psip$ data. }
\label{f_epem}
\end{figure*}

In an $e^+e^-$ collider, a virtual photon can be emitted from each lepton. The interaction of these two virtual photons will produce even $C$-parity  states such as pseudoscalar mesons, called two-photon process~\cite{twophoton}. In the case of $\etap$ production, the two-photon process $e^+e^- \to e^+e^-\etap$ leads to the same final state as signal if the outgoing $e^+$ and $e^-$ are both detected. It also contributes as a peaking background on the $M(\gamma \pi^+\pi^-)$ and $M(\gamma \gamma \pi^+\pi^-)$ distributions.  An independent $\psi(3770)$ data sample  taken at c.m. energy of $3.773$ GeV, corresponding to an integrated luminosity of 2.93 fb$^{-1}$~\cite{lumi3770,pipics}, is used to study this  background. Scatter plots of the polar angle $\cos \theta$ of $e^+$ and $e^-$ for the selected events from the signal MC sample and $\psi(3770)$ data, dominated by two-photon events, are shown in Fig.~\ref{f_epem}~(a). For the signal events, in which the electron is mostly close to the positron in direction, they mainly accumulate in the diagonal band $\cos \theta(e^+)=\cos\theta(e^-)$ in the scatter plot. For the two photon background evens, the outgoing direction of the $e^{\pm}$  approaches its ingoing beam direction thus they mainly accumulate in the bands of  $\cos \theta(e^+) >0.8$ or $\cos \theta(e^-) < -0.8$, especially in the intersection part. The corresponding scatter plot of events from $\psip$ data is shown in  Fig.~\ref{f_epem}~(b). To suppress the background from two-photon process,  $\cos \theta(e^+)<0.8$ and $\cos \theta(e^-) > -0.8$ are further required. To estimate the number of reaming two-photon background events in the $\psip$ data,  we use $\psi(3770)$ data as a normalization. After applying all above selection criteria, the number of  survived two-photon events in $\psi(3770)$ data is obtained by fitting the $M(\gamma \pi^+\pi^-)$ and $M(\gamma \gamma \pi^+\pi^-)$ distributions. A scale factor $f$ is defined as the ratio of the observed number of two-photon events $N$ in $\psip$ data to that in the $\psi(3770)$ data

$$ f \equiv \frac{N_{\psip}}{N_{\psi(3770)}} = \frac{\mathcal{L}_{\psip}}{\mathcal{L}_{\psi(3770)}}  \cdot \frac{\sigma_{\psip}}{\sigma_{\psi(3770)}}  \cdot \frac{\varepsilon_{\psip}}{\varepsilon_{\psi(3770)}} ,$$
where $N$, $\mathcal{L}$, $\sigma$ and $\varepsilon$ refer to the observed number of two-photon events, integrated luminosity of data samples ($\mathcal{L}_{\psip} = 668.55~{\rm pb}^{-1}$~\cite{tnum}, $\mathcal{L}_{\psi(3770)} = 2.93~{\rm fb}^{-1}$), cross section and detection efficiency  of two-photon process  at the two c.m. energies. The details on the cross-section can be found in Ref.~\cite{twophoton}. The detection efficiency ratios $\frac{\varepsilon_{\psip}}{\varepsilon_{\psi(3770)}}$ are determined  to be $1.10\pm0.01$ and $1.19\pm0.02$ for the two modes by the simulation with generator {\sc ekhara}~\cite{ekhara1,ekhara2}.  The scale factor is calculated to be 0.245 (0.265)  and the normalized number of the remaining two-photon background events in the $\psi(3686)$ data is  $1.4\pm1.7 $ ($0.5 \pm 0.4$) for the decay mode  $\etap \to \gamma \pi^+\pi^-$ ($\etap \to \pi^+\pi^- \eta$).

\begin{figure*}[htb]
\centering
\includegraphics[width=0.24\textwidth]{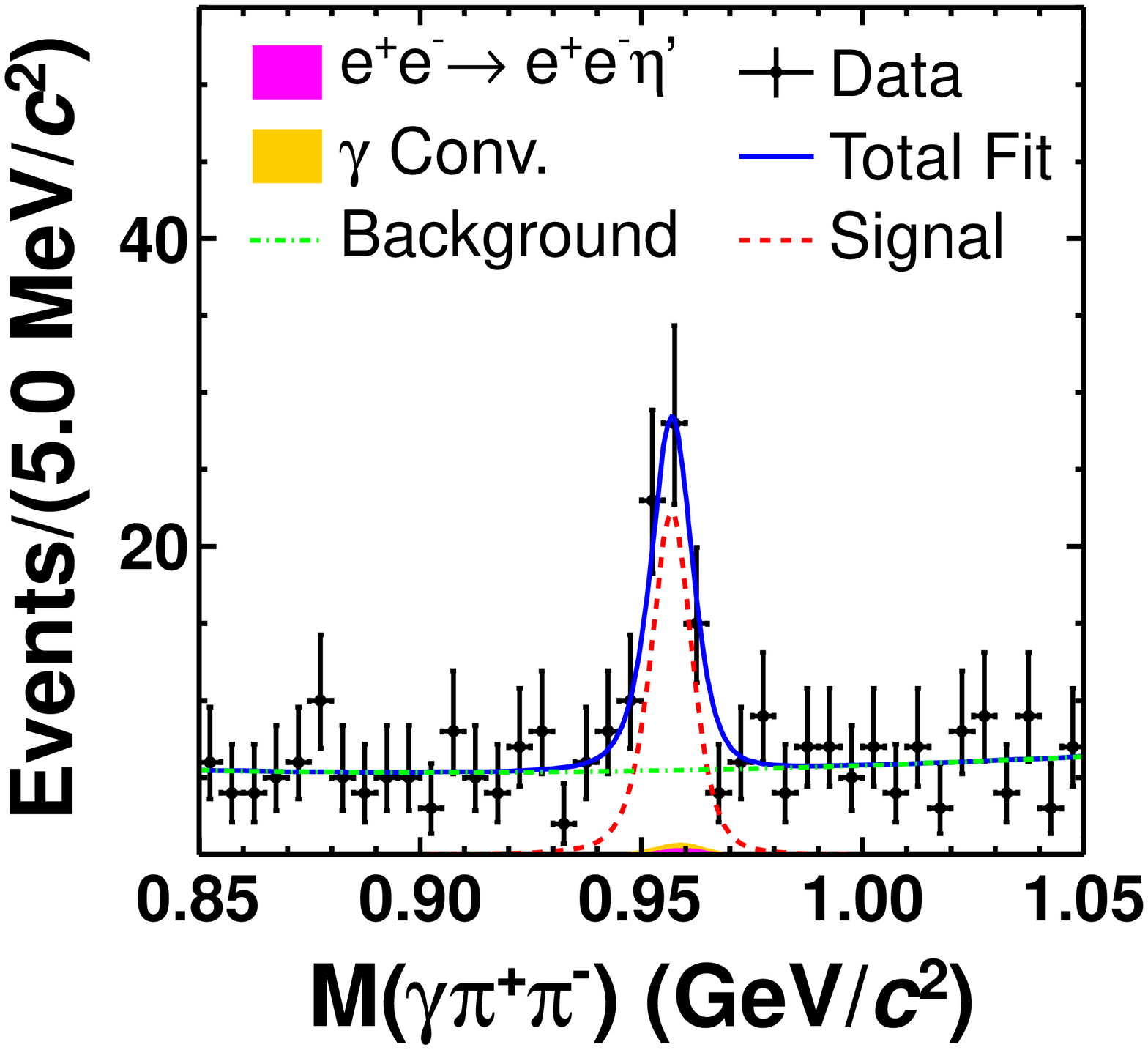}\put(-35,65){\bf   ~(a)}
\includegraphics[width=0.24\textwidth]{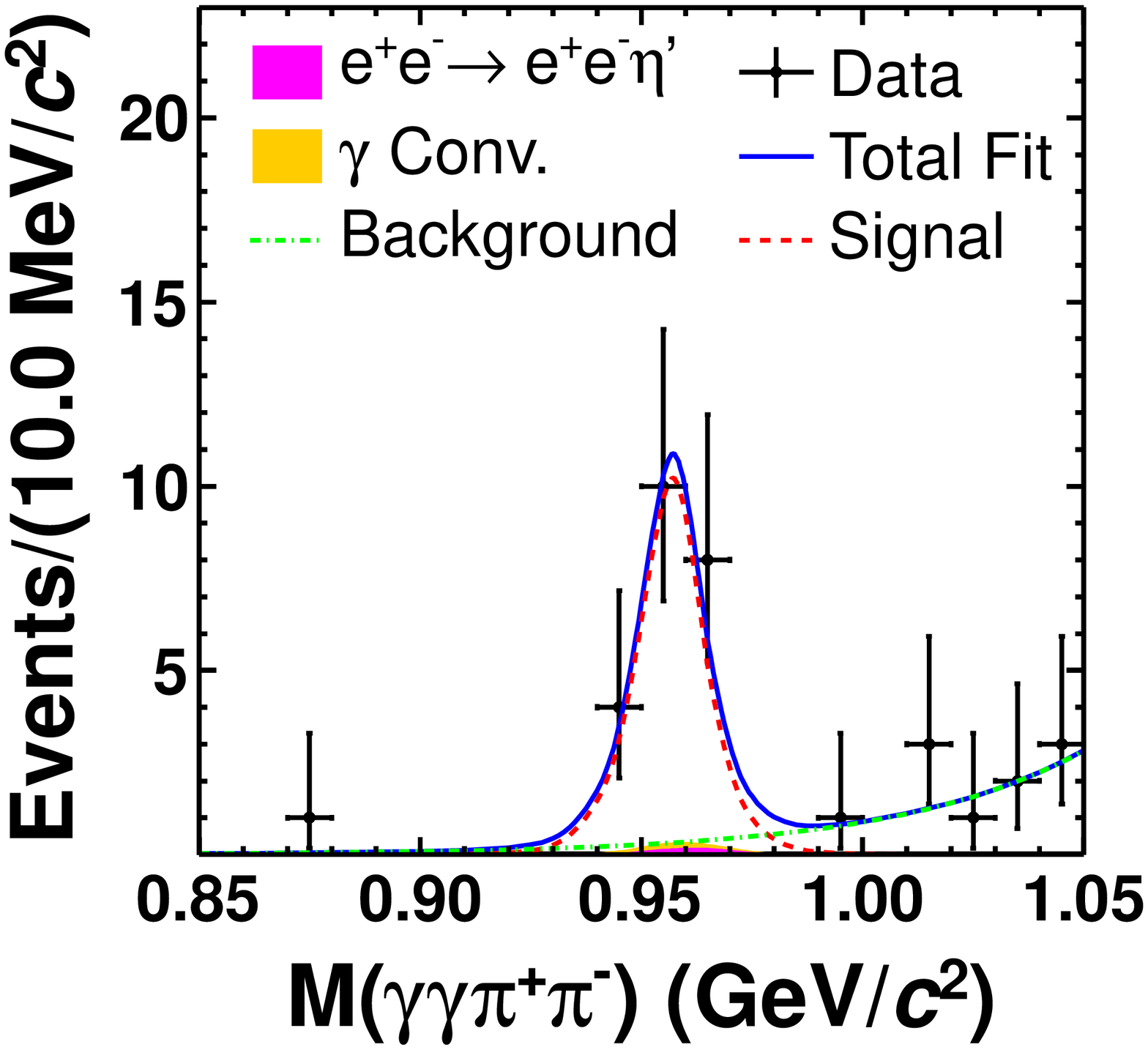}\put(-35,65){\bf   ~(b)}
\\
\includegraphics[width=0.24\textwidth]{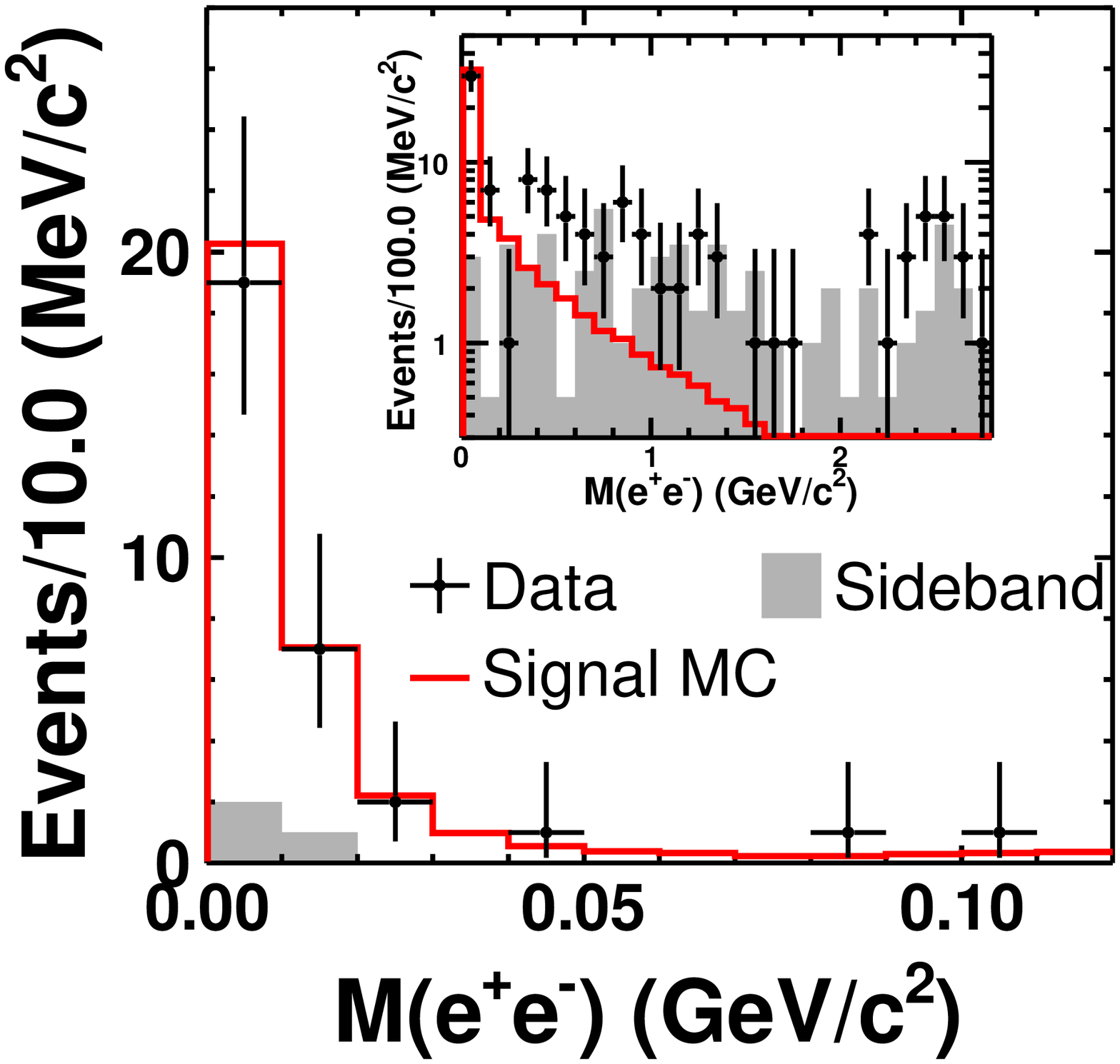}\put(-35,65){\bf  ~(c)}
\includegraphics[width=0.24\textwidth]{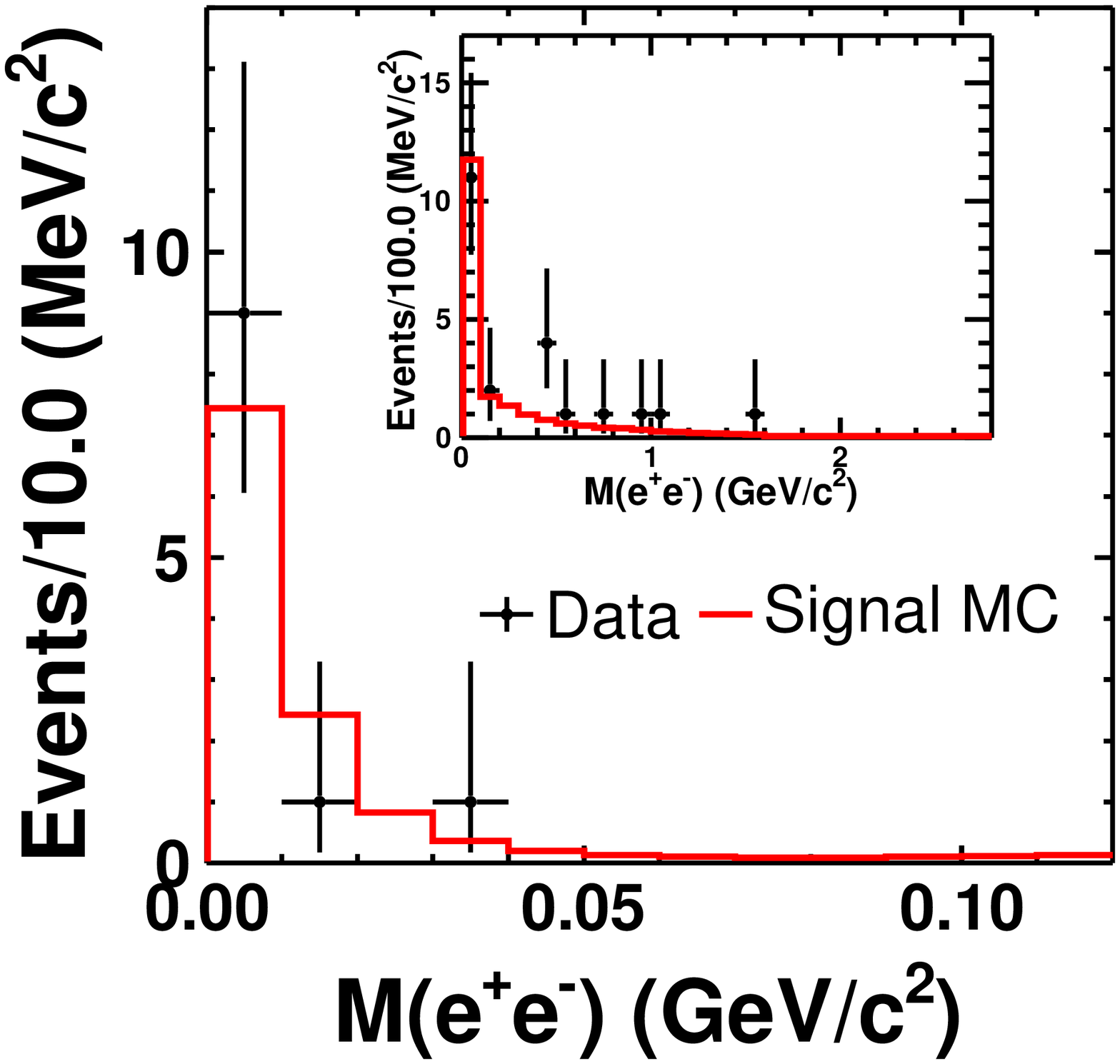}\put(-35,65){\bf  ~(d)}
\caption{(Color online.) (a, b) Mass distributions for the $\etap$ signal, (c, d) the $M(e^+e^-)$ distribution  in $\etap \to \gamma \pi^+\pi^-$ / $\etap \to \pi^+\pi^- \eta$ mode.  In (a) and (b), the black dots with error bars represent data,  the blue solid line is the total fit result, the red dashed line shows the signal,  the green dot-dashed line denotes the non-peaking background, the pink and green shaded areas indicate the peaking background from two-photon  and  $\gamma$ conversion, respectively. In (c) and (d), the black dots with error bars represent data, the red solid and  gray shaded histograms  represent signal MC simulation and non-peaking background estimated from $\eta'$ sideband, respectively, the insets show the $M(e^+e^-)$ distributions in a wider range.  }
\label{f_fit}
\end{figure*}

After applying the  above selection criteria, the studies with the inclusive MC sample indicate that the remaining background mainly arises from  $\psip \to \pi^+\pi^- J/\psi, J/\psi \to e^+e^- (\gamma)$ events, which  contributes as a non-peaking background on the $M(\gamma \pi^+\pi^-)$ and $M(\gamma \gamma \pi^+\pi^-)$ distributions.   To determine the signal yield of  $\psip \to \etap e^+e^-$, an unbinned maximum likelihood (ML) fit is performed to the $M(\gamma \pi^+\pi^-)$  and  $M(\gamma \gamma\pi^+\pi^-)$ distributions in the range of [0.85, 1.05] GeV/$c^2$, as shown in Figs.~\ref{f_fit}~(a) and~\ref{f_fit}~(b). In the fit, the signal probability density function (PDF) is described by the signal MC shape convolved with a Gaussian function, which is used to compensate  the resolution difference between data and MC simulation. The non-peaking background PDF is parameterized  with a second order Chebychev polynomial function  for the decay mode $\etap \to \gamma \pi^+\pi^-$ and with an exponential function for the $\etap \to \pi^+\pi^- \eta$ mode. The shape of the peaking background from  $\psip\to \eta' \gamma$ due to $\gamma$ conversion is derived from the MC simulation, and its magnitude is fixed to the value estimated  by taking into account the corresponding branching fractions from PDG~\cite{pdg16}. The peaking background from the two-photon process $e^+e^- \to e^+e^-\etap$  is described using the shape obtained from  $\psi(3770)$ data  and its magnitude is  fixed at  evaluated values. The corresponding distributions of $e^+e^-$ invariant mass $M(e^+e^-)$ for the candidate events within $\etap$ mass region [0.93, 0.98] GeV/$c^2$ are shown in Figs.~\ref{f_fit}~(c) and~\ref{f_fit}~(d), where the number of signal MC events is normalized to the corresponding fitted yield. The signal MC sample generated with monopole TFF  agrees well with $\psip$ data.

\begin{table*}[htb]
\caption{Signal and background yields, detection efficiency, significance  and obtained branching fraction $\mathcal{B}$  of $\psip \to \etap e^+e^-$ for $\eta' \to \gamma \pi^+\pi^-$  and  $\eta' \to \pi^+\pi^- \eta$ modes. The first uncertainties of branching fractions are statistical while the second ones are systematic.}
\begin{center}
\begin{tabular}{c c c}
\hline
\hline
 & $\eta' \to \gamma \pi^+\pi^-$  & $\eta' \to \pi^+\pi^- \eta$  \\
\hline
Signal yield &  57.4 $\pm$ 9.6 & 20.2 $\pm$ 4.3 \\
Background yield & 224.1 $\pm$ 16.2 & 12.0 $\pm$ 3.6 \\
$\epsilon$ (\%) & 22.04 & 14.89 \\
Significance  ($\sigma$) & 7.0  & 6.3 \\
$\mathcal{B}$ ($\times 10^{-6}$) & $ 1.99 \pm 0.33 \pm 0.12$ & $1.79 \pm 0.38 \pm 0.11$ \\
\hline
\hline
\end{tabular}
\end{center}
\label{t_fit}
\end{table*}

The individual branching fractions for the two $\etap$ decay modes are calculated with

$$ \mathcal{B}(\psip \to \etap e^+e^-) =\frac{N_{\rm sig}}{N_{\psi(3686)} \cdot \mathcal{B}(\eta' \to   X  ) \cdot  \epsilon},$$
where $N_{\rm sig}$  is the signal yield obtained from fitting, $N_{\psip }=(448.1 \pm 2.9) \times 10^6$~\cite{tnum} is the total number of $\psip$ events, $\mathcal{B}(\eta' \to  X)$ is the branching fraction of $\etap$ meson decaying to specific final state $X$ and quoted from PDG~\cite{pdg16}, $\epsilon$ is the detection efficiency from signal MC simulation. The statistical significance, as determined by the ratio of maximum likelihood value and that with signal contribution set to zero, are 7.0$\sigma$ and 6.3$\sigma$ for the $\etap\to \gamma\pi^+\pi^-$ and $\etap\to \pi^+\pi^-\eta$ modes, respectively. The yields obtained from the fit, the detection efficiency, statistical significance, and the obtained branching fractions for each mode are listed in Table~\ref{t_fit}, individually.

\section{\boldmath Systematic uncertainties}
Systematic uncertainties in the branching fraction measurement  are summarized in Table~\ref{t_sys}. Most of them are determined by comparing the selection efficiency of control samples between data and MC simulations.

The tracking efficiency  difference  between  data and MC simulation, both for electrons~\cite{eechicj} and charged pions~\cite{pisys}, is estimated to be 1\% for each charged track, which results in a total systematic uncertainty 4\%  for both modes.

The uncertainty associated with the photon detection efficiency, derived from a control sample of $\jpsi \to \pi^+\pi^-\pi^0, \pi^0 \to \gamma \gamma$, is 1.5\% for each photon in the endcap region and 0.5\% for each photon in barrel region. The average value, weighted according to the ratio of numbers of photon in the endcap and barrel regions, is 0.6\% for each photon. As a result, 0.6\% is assigned as the photon uncertainty in $\etap \to \gamma \pi^+ \pi^-$ mode and 1.2\% in $\etap \to \pi^+ \pi^-\eta$ mode.

The uncertainty on electron identification is studied with the control sample of radiative Bhabha scattering events $e^+e^- \to \gamma e^+e^-$. The average efficiency difference for electron identification between the data and MC simulation, weighted according to the polar angle and momentum distribution of signal MC samples, is determined to be 0.3\% for electron and positron, individually. The average efficiency difference between data and MC simulation associated with   the requirement $E/p>0.8$ is estimated to be 0.2\% with a similar method.

The systematic uncertainty related with the $\gamma$ conversion veto criterion $\delta_{xy}<2$ cm has been investigated with a control sample of $\jpsi \to \pi^+\pi^-\pi^0, \pi^0 \to \gamma e^+e^-$. The relative difference of efficiency associated with the $\gamma$ conversion rejected  criterion between data and MC simulation is 1\%~\cite{chuxkpee}, which is taken as the systematic uncertainty.

In the 4C kinematic fit, the helix parameters of charged tracks are corrected for the signal MC samples to improve the consistency between data and MC simulation, as described in Ref.~\cite{guoyp4c}. We compare the detection efficiencies obtained with and without helix parameters correction of signal MC samples. The relative change in results, 0.8\%  for $\etap \to \gamma \pi^+\pi^-$ and  1.4\% for $\etap \to \pi^+\pi^-\eta$ modes, are taken as the systematic uncertainties associated with 4C kinematic fit.

The uncertainty for the $\eta$ reconstruction using $\gamma \gamma$ pair is 1\% based on a study of a control sample of $\jpsi \to p\bar{p}\eta$~\cite{etasys}.

The uncertainty related to the $RM(\pi^+\pi^-)$ requirement is estimated by changing the selection criteria of it from 2.90 to 2.87  GeV/$c^2$  and from 3.20 to 3.17  GeV/$c^2$ for $\etap \to \gamma \pi^+\pi^-$  and $\etap \to \pi^+\pi^-\eta$ mode, respectively. The difference of branching fractions between the resulting and nominal requirement,  0.2\% and 1.9\%,  are assigned as the systematic uncertainty for the two modes, respectively.

The nominal signal MC samples are generated based on the amplitude described in Ref.~\cite{fujinlin_modphya}, where the parameter $\Lambda$ for the  monopole form factor $F(q^2)$  is set to be 3.773~GeV/$c^2$. Following the procedure used in Ref.~\cite{chuxkpee}, we adjust the $\Lambda$ to a larger value 5.0 GeV/$c^2$ or a smaller value 3.2 GeV/$c^2$ and re-generate the alternative signal MC samples. The resultant largest efficiencies change, 0.9\% and 0.2\% for two individual $\etap$ decay modes, are regarded as systematic uncertainties associated with the uncertainty from the form factor.

In the nominal fit, an MC-based shape convolved with a Gaussian function is used to model the signal PDF. An alternative fit is performed in which the signal shape is described with the MC-simulated shape only. The changes of the signal yield result, 2.6\% and 0.5\% for the individual modes, are assigned as systematic uncertainties associated with the signal shape in the fit.

The systematic  uncertainty due to non-peaking background is estimated by varying the fit range and changing its shape. In addition to the nominal fit range [0.85, 1.05] GeV/$c^2$, two alternative ones are chosen by varying the edge of the fit range by $\pm 20$~MeV/$c^2$. A third-order Chebychev polynomial function is selected as an alternative background shape for the $\etap \to \gamma \pi^+\pi^-$ mode. For the $\etap \to \pi^+\pi^- \eta$ mode,  the MC shape of the  major non-peaking background  $\psi(3686) \to \pi^+\pi^-  J/\psi, J/\psi \to \gamma e^+e^-$ is used to model the background shape.  A series of alternative fits are performed for all possible combinations of fit ranges and modeling of non-peaking background. The resultant largest difference of signal yield with respective to the nominal values, 2.8\% and  4.5\% for each mode, are taken as the systematic uncertainties.

The uncertainty arising from peaking background due to the $\gamma$ conversion process is negligible. For the two-photon process, the uncertainty associated with the scale factor is far less than the statistical uncertainties of the background events and can be ignored. We perform a series of  alternative fits, varying the input normalized number of background events following a Gaussian function with a width of the statistical uncertainty.  The standard deviation of the signal yields from these fit results, 1.3\%  and 0.7\%, are taken as uncertainties for each mode.

The uncertainty from the total number of $\psi(3686$) events is 0.6\%~\cite{tnum} and  those of quoted branching fractions  of $\mathcal{B}(\etap \to X)$ from PDG are $1.7\%$~\cite{pdg16} for both modes.

Assuming all sources to be independent in a single mode  and adding all individual contributions in quadrature,  the total relative systematic uncertainties of the $\mathcal{B}(\psip \to \etap e^+e^-)$, are determined to be 6.2\% and 7.0\% for the two $\etap$ modes, individually.

\begin{table*}[htb]
\caption{Summary of relative systematic uncertainties of the $\mathcal{B}(\psip \to \etap e^+e^-)$ (in \%). The correlated sources between two $\etap$ reconstructed modes are denoted with asterisk.}
\begin{center}
\begin{tabular}{c c c}
\hline
\hline
Sources & $\eta' \to \gamma \pi^+\pi^-$    &  $\eta'\to\pi^+\pi^-\eta$  \\
\hline
MDC tracking *   & 4.0  & 4.0 \\
Photon detection * & 0.6 & 1.2 \\
PID *  & 0.6 &  0.6 \\
$E/p > 0.8 $ & 0.2 & -- \\
Veto of $\gamma$ conversion * &  1.0 & 1.0 \\
4C kinematic fit  &   0.8 &1.4 \\
$\eta$ reconstruction &  -- & 1.0 \\
$RM(\pi^+\pi^-)$ requirement  &0.2 & 1.9 \\
Form factor  & 0.9 & 0.2 \\
Signal shape & 2.6 &  0.5 \\
Fit range and background shape  & 2.8  & 4.5 \\
Fixed peaking background & 1.3 & 0.7   \\
Number of $\psip$ events* & 0.6  & 0.6  \\
Quoted branching fractions & 1.7 & 1.7 \\
Total & 6.2 & 7.0 \\
\hline
\hline
\end{tabular}
\end{center}
\label{t_sys}
\end{table*}

\section{\boldmath Results}
The  resulting $\mathcal{B}(\psi(3686) \to \eta' e^+e^-)$ from the two $\etap$ reconstructed modes $\etap \to \gamma \pi^+\pi^-$ and $\etap \to \pi^+\pi^- \eta$ with $\eta \to \gamma \gamma$ are  (1.99 $\pm$ 0.33 $\pm$ 0.12)$\times10^{-6}$ and  (1.79 $\pm$ 0.38 $\pm$ 0.11)$\times10^{-6}$, where the first uncertainties are statistical and second ones are systematic. The measured branching fractions from the two modes are consistent with each other within their uncertainties. Following the method described in Ref.~\cite{sunzt}, the measurements from the two modes are combined, taking into account the correlation between uncertainties among the two modes, as denoted with an asterisk in Table~\ref{t_sys}. The weighted averaged result for branching fraction of $\psi(3686) \to \eta' e^+e^-$  is calculated to be $(1.90 \pm 0.25 \pm 0.11)\times10^{-6}$, where the first uncertainty is statistical and the second is systematic.

\section{\boldmath Summary}
In summary, with a data sample of $448.1 \times 10^6$ $\psip$ events collected with the BESIII detector, we observe the charmonium  EM Dalitz decay $\psip \to \etap e^+e^- $ for the first time by reconstructing $\etap$ meson via the two decay modes $\etap \to \gamma\pi^+\pi^-$ and $\etap \to \pi^+\pi^-\eta$,  with a statistical  significance of  $7.0 \sigma$ and $6.3 \sigma$, respectively. The weighted average branching fraction of  $\psip \to \etap e^+e^-$ is measured to be $(1.90 \pm 0.25 \pm 0.11)\times10^{-6}$, where the first uncertainty is statistical and second one is systematic. The observation of this process provides new information for the interaction of charmonium states with the EM field, although the statistics of current data does not allow for a precise TFF  measurement.

\section*{\boldmath Acknowledgments}
The BESIII collaboration thanks the staff of BEPCII and the IHEP computing center for their strong support. This work is supported in part by National Key Basic Research Program of China under Contract No. 2015CB856700; National Natural Science Foundation of China (NSFC) under Contracts Nos. 11235011, 11335008, 11425524, 11625523, 11635010; the Chinese Academy of Sciences (CAS) Large-Scale Scientific Facility Program; the CAS Center for Excellence in Particle Physics (CCEPP); Joint Large-Scale Scientific Facility Funds of the NSFC and CAS under Contracts Nos. U1232105, U1332201, U1532257, U1532258; CAS Key Research Program of Frontier Sciences under Contracts Nos. QYZDJ-SSW-SLH003, QYZDJ-SSW-SLH040; 100 Talents Program of CAS; National 1000 Talents Program of China; INPAC and Shanghai Key Laboratory for Particle Physics and Cosmology; German Research Foundation DFG under Contracts Nos. Collaborative Research Center CRC 1044, FOR 2359; Istituto Nazionale di Fisica Nucleare, Italy; Koninklijke Nederlandse Akademie van Wetenschappen (KNAW) under Contract No. 530-4CDP03; Ministry of Development of Turkey under Contract No. DPT2006K-120470; National Science and Technology fund; The Swedish Research Council; U.S. Department of Energy under Contracts Nos. DE-FG02-05ER41374, DE-SC-0010118, DE-SC-0010504, DE-SC-0012069; University of Groningen (RuG) and the Helmholtzzentrum fuer Schwerionenforschung GmbH (GSI), Darmstadt; WCU Program of National Research Foundation of Korea under Contract No. R32-2008-000-10155-0.

\end{multicols}
\end{document}